# Evaluation of Arterial Signal Coordination with Commercial Connected Vehicle Data: Empirical Traffic Flow Visualization and Performance Measurement

**Shoaib Mahmud, Christopher M. Day**

Department of Civil Construction and Environmental Engineering, Iowa State University, Ames, USA
Email: smahmud@iastate.edu, cmday@iastate.edu

## Abstract

Emerging connected vehicle (CV) data sets have recently become commercially available, enabling analysts to develop a variety of powerful performance measures without deploying any field infrastructure. This paper presents several tools using CV data to evaluate traffic progression quality along a signalized corridor. These include both performance measures for high-level analysis as well as visualizations to examine details of the coordinated operation. With the use of CV data, it is possible to assess not only the movement of traffic on the corridor but also to consider its origin-destination (O-D) path through the corridor. Results for the real-world operation of an eight-intersection signalized arterial are presented. A series of high-level performance measures are used to evaluate overall performance by time of day, with differing results by metric. Next, the details of the operation are examined with the use of two visualization tools: a cyclic time-space diagram (TSD) and an empirical platoon progression diagram (PPD). Comparing flow visualizations developed with different included O-D paths reveals several features, such as the presence of secondary and tertiary platoons on certain sections that cannot be seen when only end-to-end journeys are included. In addition, speed heat maps are generated, providing both speed performance along the corridor and locations and the extent of the queue. The proposed visualization tools portray the corridor's performance holistically instead of combining individual signal performance metrics. The techniques exhibited in this study are compelling for identifying locations where engineering solutions such as access management or timing plan change are required. The recent progress in infrastructure-free sensing technology has significantly increased the scope of CV data-based traffic management systems, enhancing the significance of this study. The study demonstrates the utility of CV trajectory data for obtaining high-level details of the corridor performance as well as drilling down into the minute specifics.

## 1. Introduction

Traffic signal coordination facilitates the smooth, progressive flow of vehicle platoons along signalized corridors. Many arterial highways in the US are operated under actuated-coordinated control with time-of-day plans that implement timing plans to establish a flow pattern appropriate to expected demands by the time of day, such as the morning and afternoon peak hours, off-peak periods, etc. Good progression reduces the delay for travelers on the routes prioritized by coordination, which in many arterial highways is presumed to be the traffic that is served by the major street through movements at each intersection.

Vehicle trajectory data has been employed previously to develop performance metrics on the operation of traffic signals and surface street networks to evaluate progression [1]-[6]. Previous real-world studies used data sources available internally to commercial data providers. That data was often obtained from onboard mobile devices, such as smartphones and navigation aids. In the past few years, multiple commercial providers have begun to market disaggregate connected vehicle (CV) data obtained from auto manufacturers and other original equipment manufacturers (OEM) sources. Recent results from a pooled fund study demonstrate the utility of such data in evaluating traffic signal operations [7].

The motivation for using commercialized CV data to assess operational performance is increasing because it requires no field data collection infrastructure [7] [8] [9] [10] [11]. Vehicle trajectory data makes it possible to obtain performance measures at any location along a signalized corridor without detection [12]. The increasing availability of trajectory data from commercial CV sources enables a more accurate picture of corridor operations to be developed, which may permit evaluation not only of the assumed dominant route along the corridor but also of all other origin-destination (O-D) paths through the corridor [13]. As such data emerges, there is an opportunity to develop new tools for visualization and evaluation to support decision-making. The present paper demonstrates some uses of CV data for this purpose and uses a signalized arterial corridor in Dubuque, Iowa, as a case study. The objectives of the study are as follows:

- Develop delay and speed-based performance metrics utilizing CV based tra-

jectories;

- Assess Origin-Destination (O-D) patterns through the network and generate visualizations that reveal signal operation using a cyclic view;
- Evaluate queuing and other sources of inference with the traffic flow of the corridor.

## 2. Literature Review

Floating car studies have historically been used to evaluate progression along signalized corridors. The introduction of GPS instruments improved the data collection process but still required that the analyst drive the floating cars to collect data. Such studies are labor and equipment-intensive. More recently, advances in technology have enabled alternatives, such as measuring travel times using automated vehicle identification [14] [15], high-resolution traffic signal event data [15] [16] [17] [18], and aggregated probe vehicle data [19] [20] [21] [22] [23]. Automated traffic signal performance measures (ATSPMs) enabled by high-resolution data include visualizations such as the Purdue Coordination Diagram (PCD). The PCD shows each detected vehicle arrival relative to the local green times. The presence of platoons and their arrival time relative to green intervals can be quickly ascertained from these views, distilled into aggregated metrics, and can also support the optimization of offsets [24] [25] [26] [27] [28]. However, attaining the supporting data requires that setback detectors, data storage, and data collection equipment are available.

In the past few years, CV trajectory data have become commercially available, allowing researchers to grow infrastructure-free systems for developing performance metrics for various intersection configurations at a scale. Data and analytics vendors such as INRIX, StreetLight Data, HERE, Wejo, Waze, and RITIS have emerged in this space, forming partnerships with automakers and transportation agencies to harness data from connected vehicles and allowing users to analyze traffic flow, including travel time, speeds, and O-D patterns [29]-[34]. A significant number of current efforts regarding the performance evaluation of signalized intersections and corridors focus on developing traditional metrics using CV trajectory data.

Some recently developed vehicle trajectory-based performance metrics for individual signals include average trip time and average delay [7] [11] [35]-[42], trip counts by vehicle class [31], arrival on green (AOG) [7] [11] [12] [21] [22] [23], split failures [7] [11] [23] [40] [41] [42], and downstream blockage [11] [40] [41] [42]. Aggregation of individual signal performance along the corridor is used in this study to generate corridor-wide performance metrics. In addition to CV trajectory datasets, researchers have started to analyze CV event data, such as hard braking and acceleration data, to evaluate the offline or near real-time safety warrants at scale [43] [44] [45] [46]. Moving forward, as the geographic extent and sample size of the CV dataset increase, data applications for system management may shift significantly toward the use of CV data.

## 3. Data Description and Study Corridor

### 3.1. Review of Data Sources

For this study, CV trajectory data were obtained from a commercial vendor. The data consists of vehicle waypoint records, including a journey ID, timestamp, vehicle position (latitude and longitude), speed as measured by the vehicle, and other information. These records are able to track the movement of vehicles throughout the network, with a journey beginning when the vehicle ignition is turned on and ending when the vehicle is turned off. The waypoints are generally provided once every 3 seconds.

In this study, a commercial cloud-based server (Amazon Athena) served as the main data repository, which supports SQL-like queries for data processing and retrieval. The CV data were spatially partitioned using county borders and indexed with postal codes. Raw data for the area of interest were exported from the cloud server in CSV format. The data were then processed using R to develop the visualizations and derivative performance measures presented in this study. In the future, the techniques could be implemented on the server side.

Commercial providers obtain the data by leveraging partnerships with automakers to track vehicles. The dataset follows vehicles as they move through the road network, as well as off the road network. It is necessary to carefully select the data to remove irrelevant or potentially misleading information, such as records of vehicles sitting idly in parking lots, clipping the area of interest, traversing the corridor multiple times in the same journey, and so forth. Also, some journeys in the raw data have missing waypoints (*i.e.*, longer than the expected interval between waypoints).

For this study, a large bounding box was initially drawn around the selected corridor (US 20 in Dubuque, Iowa), which captured approximately 90,000 journeys, including 10.3 million GPS waypoints over twenty weekdays in October 2021. Later, geofences were created surrounding the corridor and portions of the side streets. Most of the irrelevant waypoints were removed by the application of the geofences. Next, journeys with fewer than five waypoints and journeys with waypoints more than ten seconds apart were removed. After these processes, the data contained 53,656 unique vehicle journeys containing approximately 3.9 million waypoints over a four-week period. **Figure 1** shows an illustration of raw vehicle waypoints for the larger bounding box and vehicle waypoints for the corridor and the side streets of the signals for 24 hours.

Next, a map-matching process was undertaken to convert the latitude and longitude data into linear distances along the corridor. First, we determined the origin-destination of each journey using geofences of the entry-exit locations of the corridors and side streets. Each waypoint was then "snapped" to a road segment using the closest distance between the waypoint and the segment. Because of the simplicity of the study corridor, a relatively simple map-matching process was sufficient. Future work will integrate more sophisticated techniques to generalize to larger-scale networks, most likely by utilizing a directed graph structure [47] [48] [49].

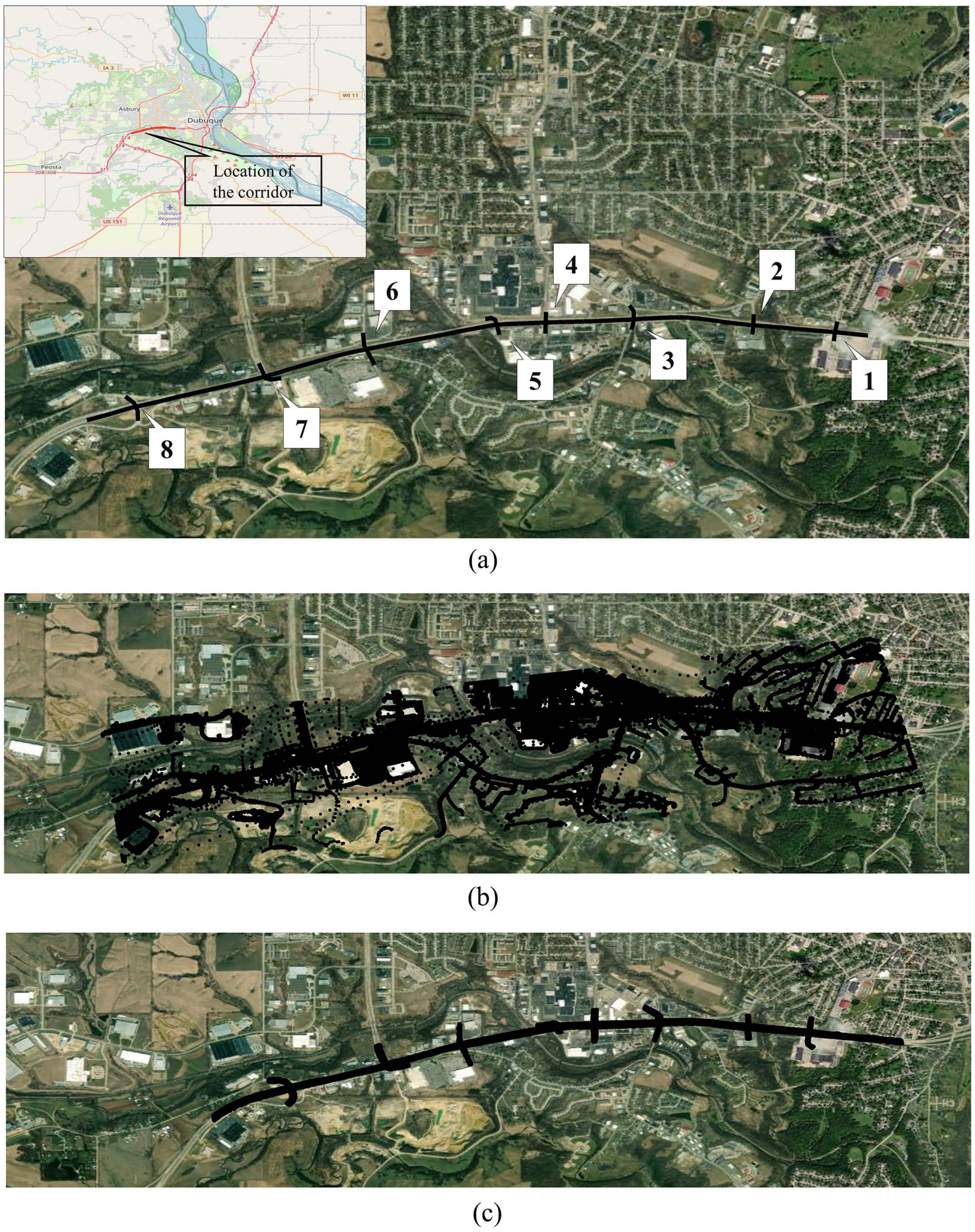

**Figure 1.** Maps of the study corridor and vehicle waypoints for 24 hours (October 4, 2021): (a) corridor location and the intersections; (b) all vehicle waypoints for the larger bounding box surrounding the study corridor (454,331 waypoints); (c) vehicle waypoints only for the corridors and the approaches of the side streets (137,215 waypoints).

After map matching, it was possible to determine the penetration rate. It was found to be approximately 3% - 6% of the annual average daily traffic (AADT). The low penetration rate was mitigated by the aggregation of data across multiple days of data, as demonstrated in previous studies [12].

In addition to trajectory data, the signal timing plan was obtained from the City of Dubuque. The weekday day plan on US 20 is as follows:

- Early morning, 06:30-08:00, Cycle Length = 109 sec;
- Morning peak, 08:00-10:30, Cycle Length = 103 sec;
- Midday, 10:30-14:30, Cycle Length = 115 sec;
- Afternoon peak, 14:30-18:00, Cycle Length = 130 sec;
- Evening, 18:00-22:00, Cycle Length = 103 sec.

Signal systems running under time of day coordination often operate the same timing plan on all weekdays. US 20 in Dubuque runs under such a scheme. This makes it possible to aggregate multiple days of data to increase the number of samples, while still capturing patterns that are consistent from day to day, as demonstrated in previous work on probe data performance measures with small sample sizes of automatic vehicle identification data [50] and detector-free offset optimization [12]. This paper adopts a similar strategy to increase the number of observations. This strategy relies on the similarity of operations from day to day.

## 3.2. Corridor Description and Origin Destination Patterns

A 2.6-mile east-west section of US 20 in Dubuque, Iowa, was selected for analysis. The corridor contains eight signalized intersections and is one of the busiest roads in that region. To understand the traffic patterns occurring on the corridor, an analysis of journeys by origin and destination was carried out. The data processing method described in the previous section resulted in every journey being associated with a particular origin and destination, making it simple to perform this analysis. Origins and destinations for the test corridor are shown in Figure 2(a), while Figure 2(b) shows an O-D matrix color coded to highlight the more dominant movements.

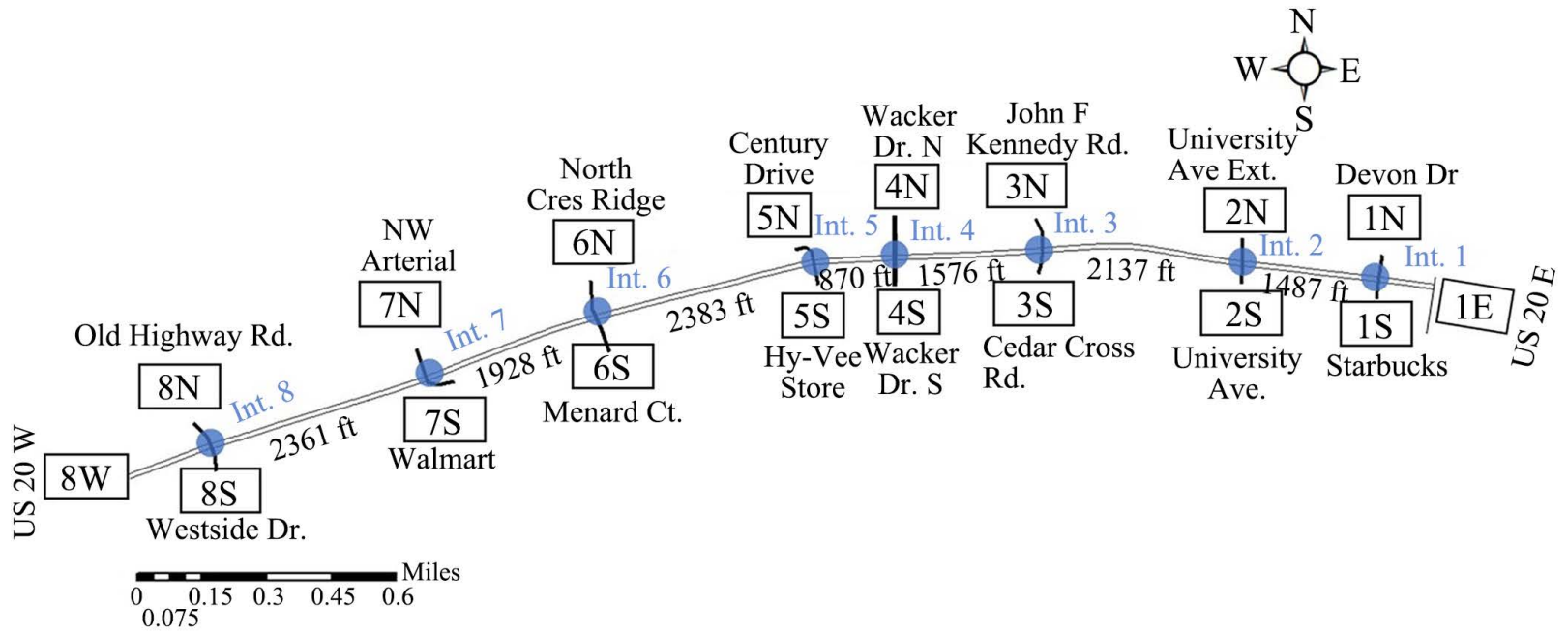

(a)

| Origin \ Destination | 1E | 1N | 1S | 2N | 2S | 3N | 3S | 4N | 4S | 5N | 5S | 6N | 6S | 7N | 7S | 8N | 8S | 8W | Total |
|---|---|---|---|---|---|---|---|---|---|---|---|---|---|---|---|---|---|---|---|
| 1E | 0 | 146 | 3 | 305 | 817 | 182 | 4136 | 4 | 1556 | 323 | 446 | 19 | 786 | 44 | 8 | 19 | 1203 | 6152 | 16149 |
| 1N | 5 | 0 | 1 | 1 | 4 | 3 | 17 | 1 | 1 | 11 | 3 | 1 | 5 | 4 | 11 | 9 | 2 | 5 | 84 |
| 1S | 123 | 3 | 0 | 1 | 3 | 2 | 16 | 4 | 13 | 3 | 5 | 1 | 1 | 5 | 2 | 17 | 5 | 48 | 252 |
| 2N | 6 | 6 | 1 | 0 | 6 | 21 | 56 | 1 | 210 | 56 | 29 | 2 | 220 | 9 | 4 | 5 | 187 | 566 | 1385 |
| 2S | 1272 | 65 | 5 | 5 | 0 | 2 | 1 | 11 | 6 | 3 | 33 | 1 | 2 | 4 | 5 | 9 | 1 | 2 | 1427 |
| 3N | 19 | 3 | 2 | 2 | 1 | 0 | 3 | 1 | 22 | 35 | 74 | 3 | 76 | 16 | 5 | 1 | 101 | 516 | 880 |
| 3S | 3343 | 430 | 1 | 20 | 63 | 4 | 0 | 5 | 5 | 5 | 17 | 1 | 0 | 0 | 2 | 1 | 4 | 3 | 3904 |
| 4N | 5 | 2 | 1 | 1 | 2 | 0 | 1 | 0 | 6 | 141 | 2 | 7 | 433 | 22 | 5 | 3 | 284 | 1270 | 2185 |
| 4S | 1547 | 223 | 1 | 185 | 28 | 1 | 38 | 0 | 0 | 1 | 2 | 0 | 1 | 5 | 4 | 0 | 0 | 6 | 2042 |
| 5N | 2 | 2 | 5 | 3 | 5 | 2 | 0 | 0 | 3 | 0 | 1 | 3 | 5 | 13 | 0 | 1 | 200 | 1124 | 1369 |
| 5S | 329 | 46 | 5 | 38 | 7 | 86 | 10 | 4 | 6 | 0 | 0 | 4 | 1 | 3 | 4 | 0 | 5 | 2 | 550 |
| 6N | 1 | 7 | 11 | 1 | 2 | 0 | 5 | 0 | 0 | 5 | 1 | 0 | 14 | 12 | 1 | 1 | 327 | 317 | 705 |
| 6S | 816 | 139 | 2 | 152 | 13 | 256 | 22 | 371 | 20 | 3 | 1 | 2 | 0 | 4 | 3 | 1 | 1 | 1 | 1807 |
| 7N | 5 | 1 | 5 | 6 | 1 | 1 | 4 | 2 | 1 | 2 | 2 | 1 | 2 | 0 | 3 | 121 | 37 | 4359 | 4553 |
| 7S | 7 | 8 | 0 | 5 | 0 | 11 | 1 | 0 | 4 | 0 | 0 | 5 | 4 | 0 | 0 | 2 | 1 | 2 | 50 |
| 8N | 1 | 11 | 1 | 8 | 1 | 2 | 9 | 3 | 0 | 1 | 4 | 5 | 2 | 72 | 5 | 0 | 1 | 51 | 177 |
| 8S | 3 | 5 | 2 | 7 | 0 | 11 | 6 | 22 | 0 | 4 | 0 | 8 | 2 | 98 | 1 | 1 | 0 | 1 | 171 |
| 8W | 6488 | 592 | 1 | 588 | 60 | 1008 | 175 | 1448 | 171 | 1553 | 1 | 787 | 19 | 2987 | 3 | 3 | 82 | 0 | 15966 |
| Total | 13972 | 1689 | 47 | 1328 | 1013 | 1592 | 4500 | 1877 | 2024 | 2146 | 621 | 850 | 1573 | 3298 | 66 | 194 | 2441 | 14425 | 53656 |

(b)

Figure 2. US-20 and trip count summary: (a) US-20 in Dubuque, IA with origins, destinations, and intersection IDs; (b) Journey count heatmap by origin-destinations with trajectory data.

We limited possible origins and destinations to roads entering or exiting the corridor and excluded driveways along the route. The O-D matrix shows that although the largest volumes are for those paths traversing the corridor end-to-end (1E-8W and 8W-1E), Cedar Cross Rd (Int. 3) and NW Arterial (Int. 7) also attract and generate a substantial of traffic, with their numbers of originating or terminating journeys approximately 1/2 to 2/3 of that of the corridor endpoints.

# 4. Methodology

## 4.1. Delay-Based Performance Measures

Travel time and delay are perhaps the most common performance measures for evaluating vehicle travel in traffic signal systems. Delay is used for intersection Level of Service (LOS) evaluations in the *Highway Capacity Manual* (HCM). For the assessment of corridor performance, the HCM LOS uses the ratio of estimated travel time to free flow travel time [51]. Such an analysis would typically only consider travel along the major through directions of the corridor. Historically it has been challenging to collect data for even those paths, let alone all of the others. However, as previous results have illustrated, many different O-D paths on a corridor may substantially affect its operation. Recent studies have shown that better consideration of O-D patterns can yield substantial improvements in signal timing [13]. A fundamental problem with comparing delay or travel time for different O-D paths is that each path has a different free flow travel time, depending on the distance and free flow speeds. One approach to resolving this is by normalizing the travel time, while another is by calculating delay.

Two possible normalized metrics are the travel rate and the travel time index (ratio of actual to ideal travel time). The travel rate $r_i$ experienced by vehicle $i$ is defined as

$$r_i = \frac{t_i}{D} \tag{1}$$

where $t_i$ is the observed travel time and $D$ is the distance traveled.

The travel time index $T_i$ is defined as

$$T_i = \frac{t_i}{t_f} \tag{2}$$

where $t_i$ is the observed travel time, and $t_f$ is the free flow travel time.

Delay $d_i$ for vehicle $i$ is defined as

$$d_i = t_i - t_f \tag{3}$$

where $t_i$ is the observed travel time and $t_f$ is the free flow travel time. In Equations (2) and (3), the speed limit is used as free flow speed for simplicity and to avoid potentially specifying an ideal travel time that may require exceeding the speed limit.

## 4.2. Smoothness of the Flow of the Traffic

Travel time and delay are single values that describe overall performance but re-veal little about what occurs during the journey. As explored in a previous study, similar travel times can result from few stops or many stops. In the past, some have tried to capture the influence of stops by adding "stop penalties" to convert stops into an equivalent travel time [3]. However, this necessitates defining what a "stop" is, and it is somewhat a matter of opinion whether a vehicle that avoids stopping by approaching an intersection slowly fares better than a vehicle that pulls up and stops, at least from the viewpoint of the driver.

In anticipation of widespread trajectory data in the future, Beak *et al.* [52] proposed a metric called the Smoothness of the Flow of Traffic (SOFT), which quantifies the platoon progression along a corridor incorporating all of the speed changes encountered by the vehicle as it traverses the corridor. The authors applied a fast Fourier transform of the frequency content of the speed data for each trajectory, yielding a score that tends to decrease as the trajectory shifts away from an ideal straight-line shape. SOFT is scaled from 0 - 100, with 100 representing ideal progression and lower values indicating poorer performance. Beak defined SOFT for a vehicle *i* as [53]

$$\text{SOFT}(i) = 100 \times \left( 1 - \sqrt{\sum_{k=1}^{N-1} \left( \frac{P_k^i}{P_0^k} \right)^2} \right) \tag{4}$$

where $P_k^i = \left| X_k^i \right|^2$, which is given by the discrete Fourier transform (33):

$$X_k^i = \sum_{n=0}^{1} s_i(nT_0) \exp\left[ \frac{-2\pi knjt}{T_0} \right] \quad \text{for } k = 0, 1, \cdots, N-1 \tag{5}$$

In the above, $s_i(nT_0)$ is the speed of vehicle *i* at measurement $nT_0$, where *n* is the measurement number and $T_0$ is the interval between measurements. Further information about the physical interpretation of SOFT is given by Beak [53]. The resulting performance measure ranges from 0 to 100, with 100 representing an ideal trajectory shape and lower values having an increasing amount of perturbations.

**Figure 3** shows example calculations of SOFT for two trajectories, illustrated using a time-space diagram and a speed-time diagram. **Figure 3(a)** shows the trajectory of a vehicle that stops once, with a high SOFT value, while **Figure 3(b)** shows a vehicle that stops multiple times and has a much lower SOFT value.

The preceding metrics are calculated for each individual vehicle trajectory, which permits the development of distributions and aggregated values for additional assessment. For delay, we can simply take the total delay recorded by all vehicles on all the O-D paths to be included in the analysis:

$$X = \sum_{j \in Z} \left( \frac{1}{n_j} \sum_{i=1}^{n_j} d_i \right) \tag{6}$$

Here $n_j$ is the number of observations for OD path *j*, and the summation is carried out for all the O-D paths in set *Z*. Sets of O-D paths could be selected to include only routes in one direction or only end-to-end movements and other combinations desired by the analyst. For the other performance measures, we take the weighted average:

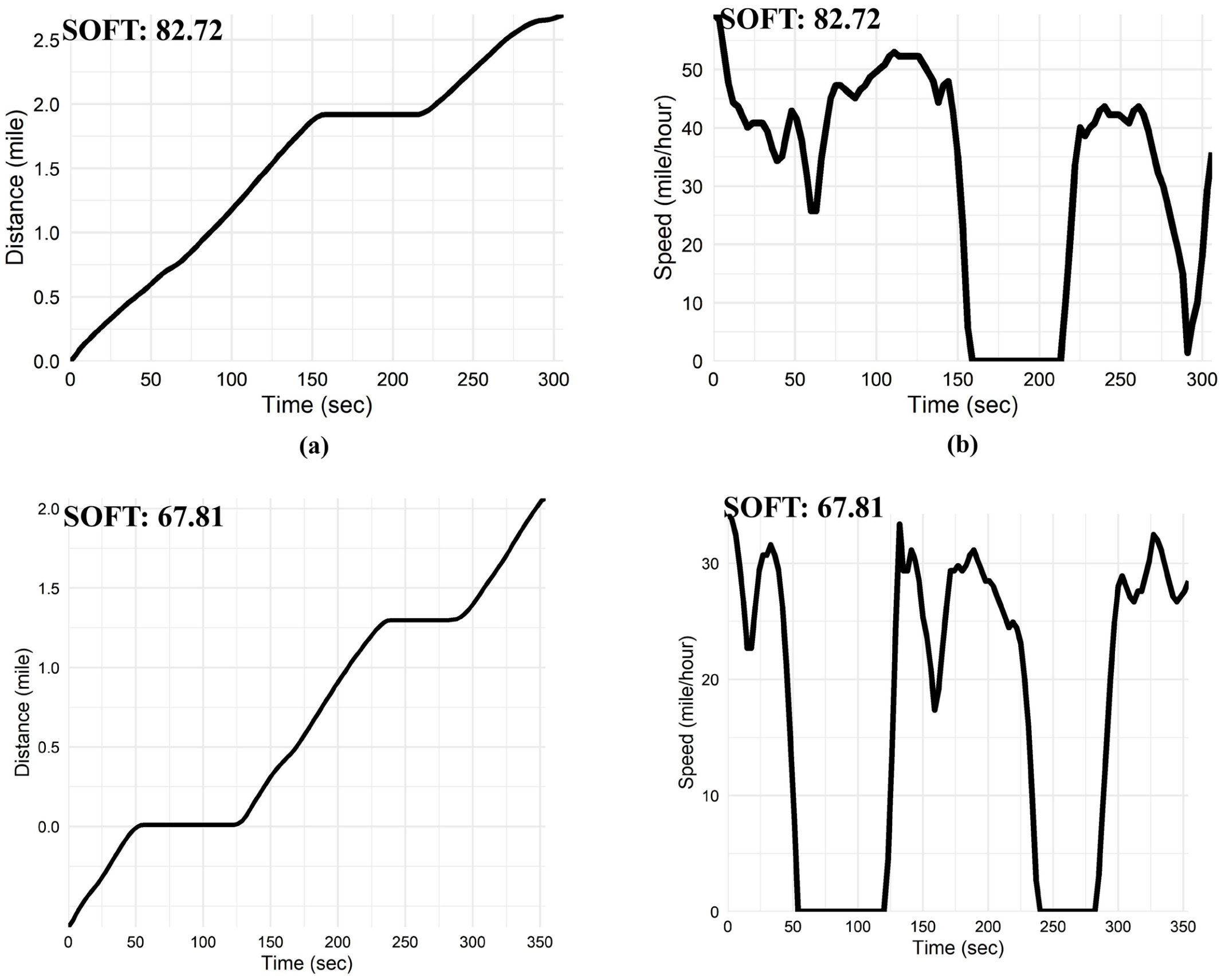

**Figure 3.** Trajectory and speed profile: (a) a vehicle experiencing few stops and slowdowns; (b) a vehicle that experiencing many stops and slowdowns.

$$X = \frac{\sum_{j \in Z} w_j \bar{x}_j}{\sum_{j \in Z} w_j} = \frac{\sum_{j \in Z} w_j \left( \frac{1}{n_j} \sum_{i=1}^{n_j} x_i \right)}{\sum_{j \in Z} w_j} \quad (7)$$

Here, $X$ is the final aggregate performance measure, $w_j$ is a weight value for O-D path $j$, $\bar{x}_j$ is the average for O-D path $j$, $n_j$ is the number of observations for O-D path $j$, and $x_i$ is the $i$th observation. The weight value used for each O-D path was the number of journeys observed on that path during the period of interest.

## 4.3. Corridor Level Performance Visualization

In addition to quantitative performance measures, spatiotemporal visualizations are useful for transportation agencies to assess the system-level operation and to discover the locations of problems and their nature. A time-space diagram (TSD) is one option and is particularly useful for viewing vehicle trajectories. Another visualization that may be particularly useful is the "platoon progression diagram" (PPD) proposed in 1984 by Wallace and Courage [54]. Historically in practice, such diagrams have only been available using the signal timing data with modeled traffic flows, which may vary from the real traffic flow because of variations in the volume and actuation. This study uses CV data to demonstrate the construction of empirical, cyclic TSDs and PPDs.

The PPD is similar in concept to cyclic flow profiles, which were employed in the seminal work on flow-based optimization of traffic signal offsets [55], notably in the widely used signal timing software TRANSYT [56]. The PPD is an extension of the flow profile into a spatial dimension. The PPD is available as a visualization in more recent UK versions of TRANSYT and in the software TruTraffic. Since the PPD considers a spatial dimension, it could also potentially be regarded as a cyclic flow and density diagram.

In this paper, to deal with the relatively low sample rate, the TSD, and PPD are presented as cyclic aggregations, which require that time of day be transformed to time in the cycle. The principle behind this is derived from how most signal controllers determine their local cycle time during coordination. For any given timestamp, the time of day $t$ can be stated in terms of the number of seconds after a daily reference point. Most controllers use midnight as their default daily reference point. The time in the cycle for an event occurring at $t$ occurs at the following time in cycle ($\tau$)

$$\tau = t \bmod C \tag{8}$$

where $C$ is the cycle length (s) in effect at time $t$ (s) after midnight.

In graphical terms, the above transformation effectively truncates our TSD when the time axis reaches $C$, and causes the trajectories to "wrap around" to the other side. The distance axis does not change. Whereas the resulting visualization no longer represents the condition of any particular cycle, the overlay of many trajectories in this view offers a useful way to assess locations where vehicles stop and slow down as they traverse the corridor. Furthermore, it is possible to select which O-D paths are included in the diagrams: only vehicles traveling from end to end, or those that only partially traverse the corridor.

An empirical PPD can show similar data as two-dimensional distributions rather than overlaid trajectories. In a PPD, the number of vehicles observed can be visualized as a two-dimensional bin in a discretized time-space field, becoming a pixel in the overall diagram. To create an empirical PPD, we begin by choosing time interval $\Delta t$ and space interval $\Delta x$ to discretize the relevant axes. For this study, we used 1-second bins for $\Delta t$ and 100 ft divisions for $\Delta x$. Exploiting these divisions, it is possible to locate the address of any pixel in ($x$, $t$) coordinates where $t = [0, C-1]$ and $x = [0, N_x]$, where $N_x$ is the number of spatial divisions given by

$$N_x = \text{floor}\left[\frac{L}{\Delta x}\right] \tag{9}$$

The "floor" function used above deletes the non-integer portion of the contained elements.

Figure 4 illustrates the construction of the basic elements of the cyclic TSD and the empirical PPD. Figure 4(a) shows two diagrams in which the trajectories are transformed from "real-time scale" to "time in the cycle scale" using Equation (8). The method utilized in earlier studies for generating a cyclic arrival profile or creating a cyclic time-space diagram involves aggregating trajectories with a low sampling rate [12] [55] [57]. The incorrect cycle length has been used in the left side figure, and the data are overlaid with no clear pattern. This purposefully illustrates the effect of even the slightest error in cycle length. The right-side figure uses the correct cycle length, and cyclic flow patterns become apparent. Vehicles tend to stop at the first intersection, many of them stop at the second, and there are significant new vehicles entering the corridor after the second intersection. One can visually traverse the intersection and characterize patterns in traffic flow. Figure 4(b) shows the further transformation of a few trajectories into the type of data representation for the PPD. Three trajectories are initially shown in a discretized time-space field. Next, for each pixel in that field, we count the number of trajectories that make an appearance within the relevant temporal and spatial divisions and then color the pixel in proportion to the total count.

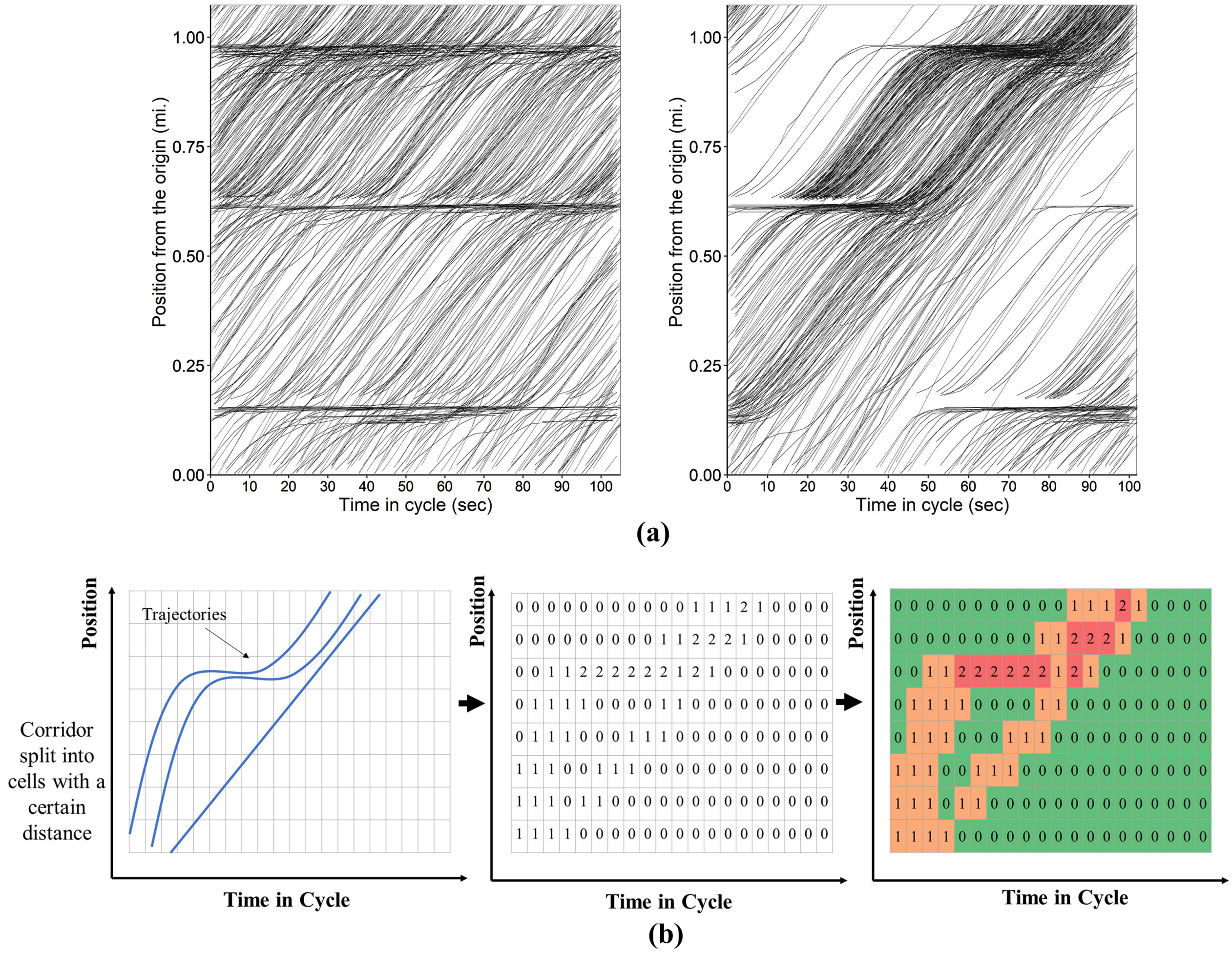

Figure 4. Development of empirical platoon progression diagram: (a) projection of trajectories into one representative cycle with incorrect cycle length vs correct cycle length; (b) converting time-space diagram to platoon progression diagram.

CV data enables these visualizations to go further than previous iterations of the TSD and PPD since the O-D path associated with each trajectory is known. Each of these visualizations can be done for any given O-D path in the network and for selected O-D paths inside of that particular view. These visualizations permit the analyst to "drill down" from high-level aggregate values of the type described in the previous section to examine the details of the operation.

### 4.4. Corridor Level Speed Heat Map and Queue Visualization

The presence of queues and other sources of interference with traffic flows are important to manage traffic signal timing in corridors. To examine queuing and other disruptions, we propose a speed heat map that illustrates where such events occur and how prevalent they are throughout a typical day. For the speed heat map, we use 1-hour bins for $\Delta t$ and the previously defined 100 ft divisions for $\Delta x$. Now, for any pixel in ($x$, $t$) coordinates, where $t = [0, 23]$ and $x = [0, N_x]$ the average of the measured speed of all vehicle waypoints in that cell can be given by $\bar{s}_{t,x}$

$$\bar{s}_{t,x} = \frac{\sum_{0}^{n_{t,x}} s_{t,x}}{n_{t,x}} \tag{10}$$

where $n_{t,x}$ is the number of observations in the pixel.

When traffic is queued, the average speed associated with a pixel should be much lower compared to the pixels with normal traffic conditions. The queued state can be identified by choosing a speed threshold. Several ways exist to select a speed threshold; however, the speed limit and free flow speed have often been used as a reference [58] [59]. In the present study, the posted speed limit along the corridor is 45 mph. We selected 35 mph speed as the threshold for indicating queued conditions and color-coded the graphics accordingly.

## 5. Results and Discussion

Using the emerging CV-based trajectory datasets, performance metrics were developed to assess the quality of traffic progression. These include total delay, travel rate, travel time index, and SOFT, which collectively provide a comprehensive overview of corridor performance (**Figure 5**). Additionally, three visualizations were introduced—cyclic time-space diagram (TSD) (**Figure 6**), empirical platoon progression diagram (PPD) (**Figure 7** and **Figure 8**), and corridor speed heat maps (**Figure 9**)—to examine both high-level trends and intricate details of corridor performance. The performance metrics generated are examined and discussed, with an emphasis on interpreting the outcomes and delineating various facets of the resulting visualizations.

### 5.1. Travel Time and Delay-Based Performance Metrics

The weighted average method described in Equation (7) was used to combine

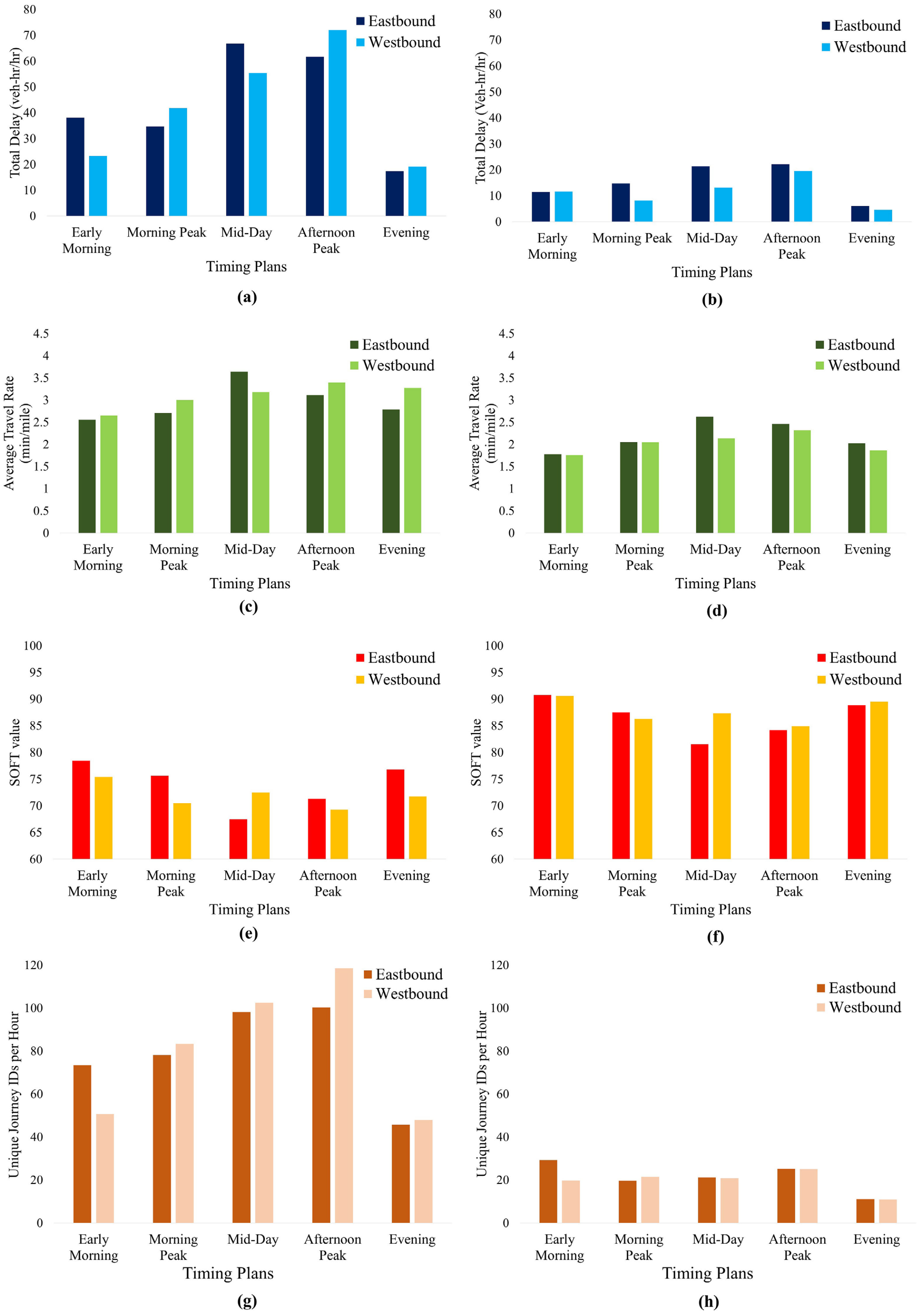

**Figure 5.** Comparison of the performance measure metrics: (a) composite delay for all routes; (b) composite delay for end-to-end journeys; (c) average travel rate for all routes; (d) average travel rate for end-to-end journeys; (e) average SOFT for all routes; (f) average SOFT for end-to-end journeys; (g) unique journey IDs per hour for all routes; (h) unique end-to-end journey IDs.

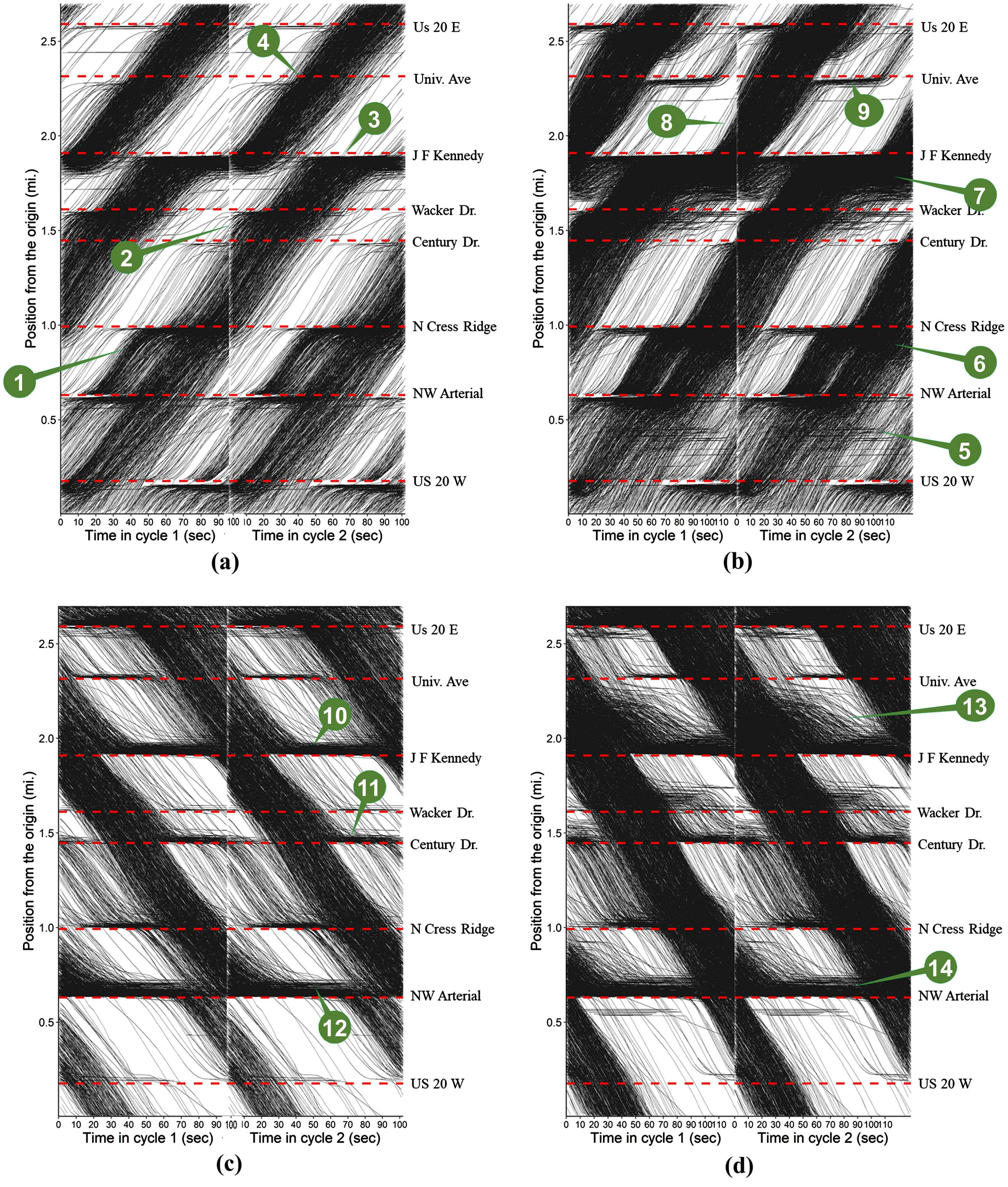

Figure 6. Cyclic TSDs for end-to-end vehicle trajectories: (a) eastbound/morning peak; (b) eastbound/afternoon peak; (c) westbound/morning peak; (d) westbound/afternoon peak.

the values of the quantitative performance measures presented earlier. This was done for each time of day plan. The outcomes of this process are illustrated in Figure 5. Some selected observations are made below:

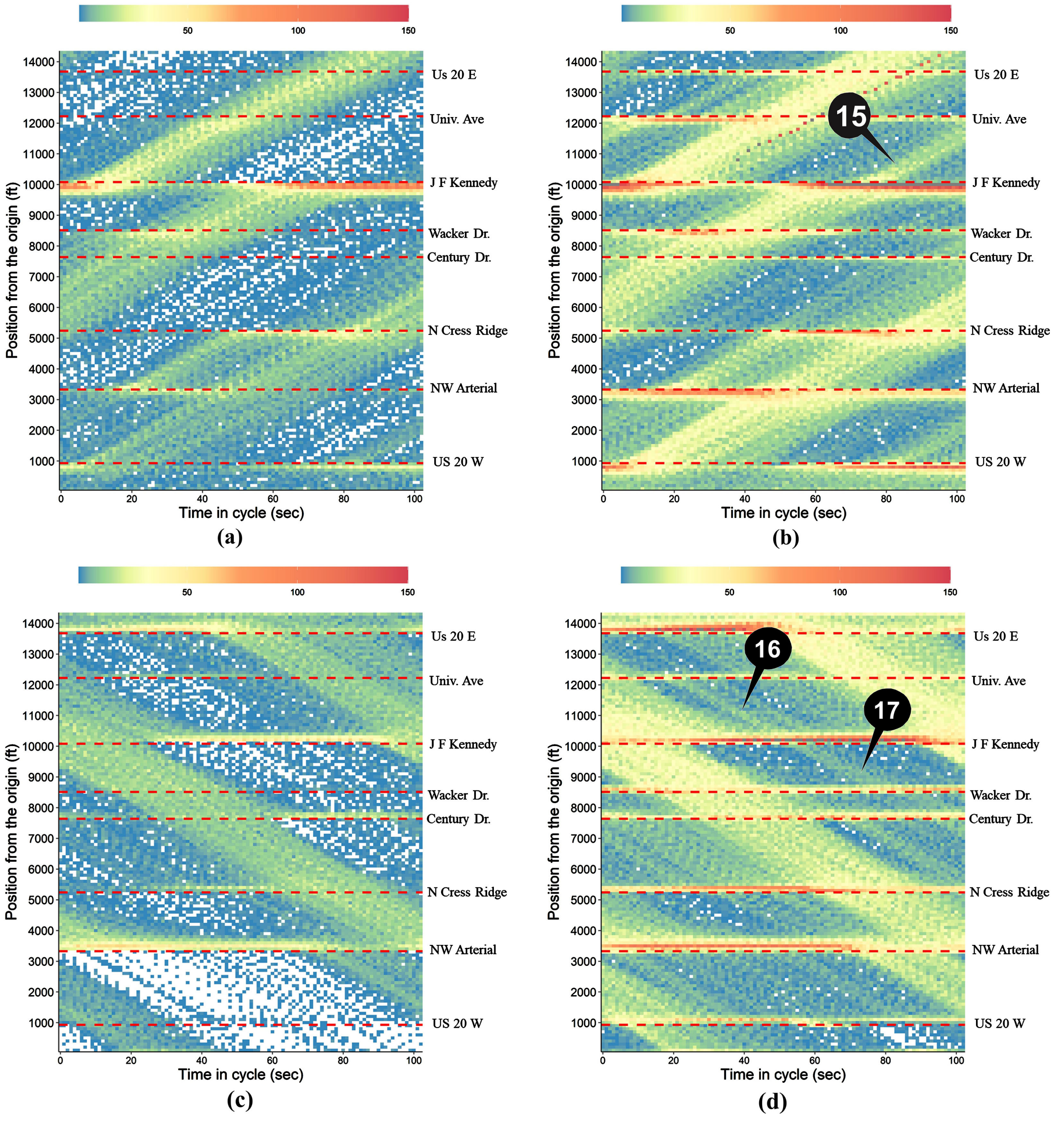

Figure 7. Empirical PPDs during the morning peak: (a) eastbound, end-to-end journeys; (b) eastbound, all journeys; (c) westbound, end-to-end journeys; (d) westbound, all journeys.

- For total delay, when considering all journeys (Figure 5(a)), westbound/ afternoon peak exhibits the worst performance, followed by eastbound/ midday. For end-to-end journeys (Figure 5(b)), the eastbound direction appears worse during both times of day. Comparing Figure 5(b) with Figure 5(a) shows that the total delay incurred by end-to-end journeys is much smaller than that of all journeys. The evening has the least delay, which is attributable to lower traffic volume during that time period. Unexpectedly, the total delay for all journeys is higher for eastbound/early morning than in the same direction for the morning peak. This is likely because the travel demand in the eastbound early morning is similar to the morning peak (Figure 5(g), Figure 5(h)), yet a different timing plan is in effect, which may not serve the eastbound traffic as well as the morning peak timing plan.

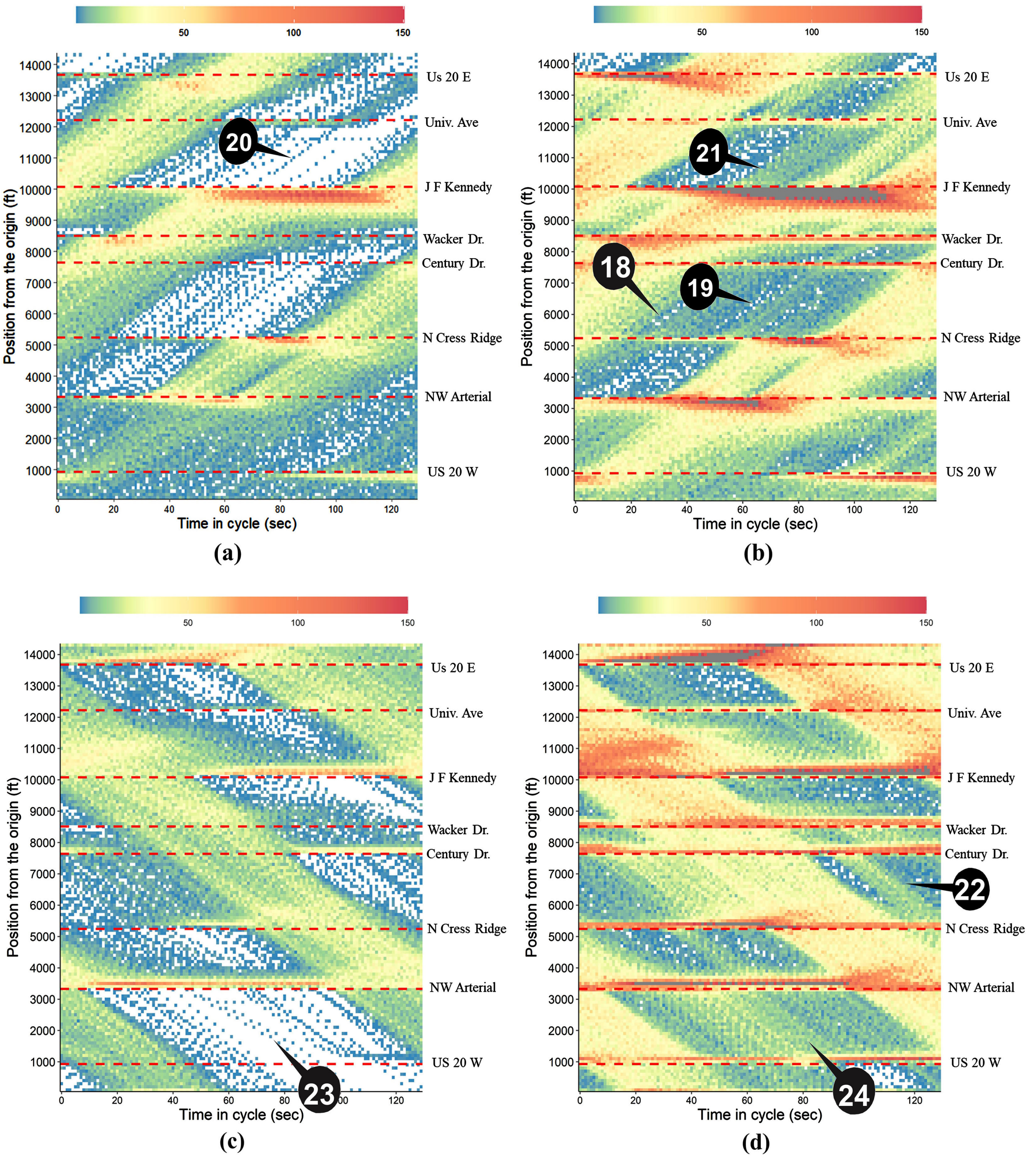

**Figure 8.** Empirical platoon progression diagrams during afternoon peak: (a) eastbound, end-to-end journeys; (b) eastbound, all journeys; (c) westbound, end-to-end journeys; (d) westbound, all journeys.

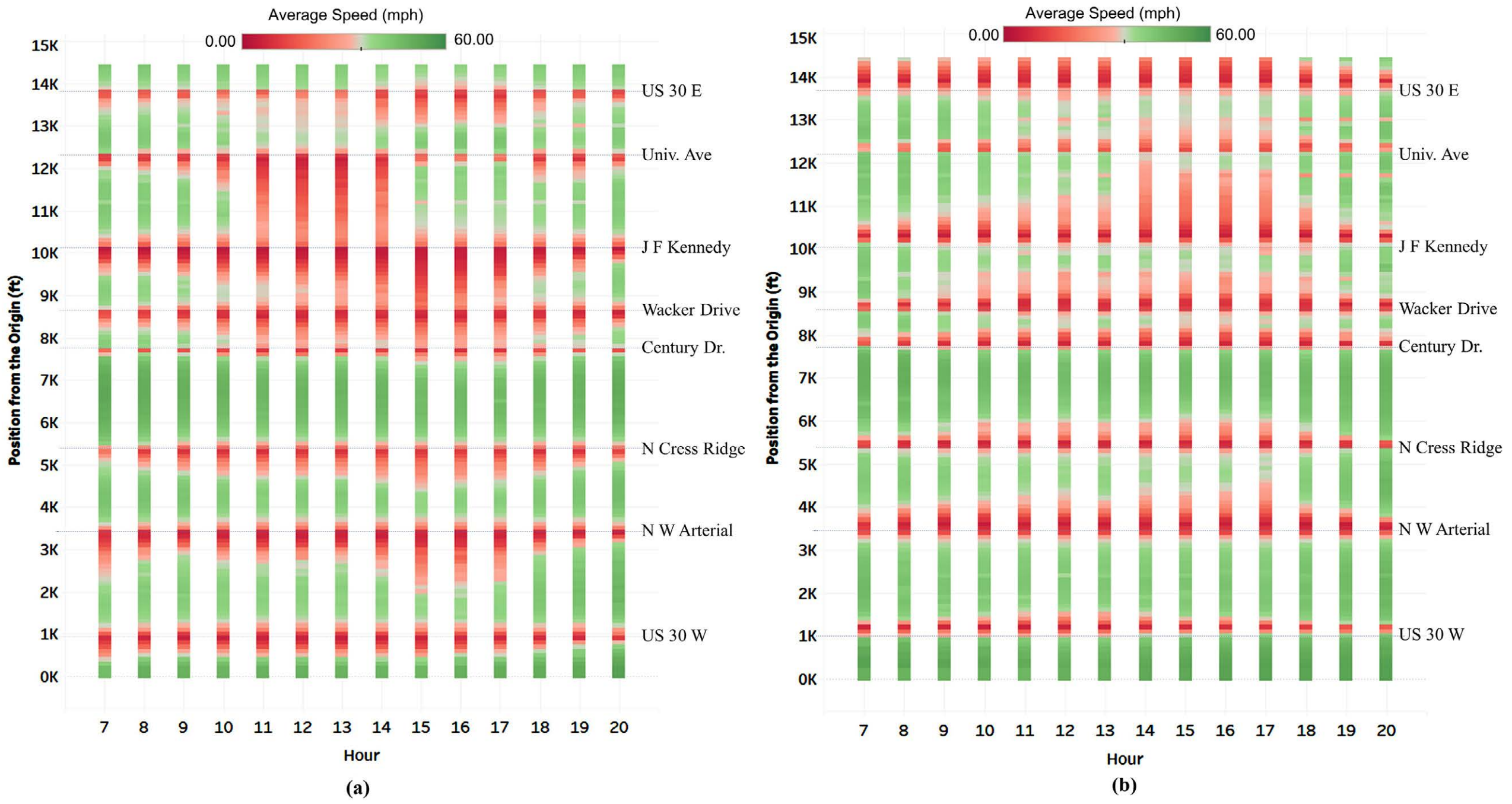

**Figure 9.** Speed heat maps by hour: (a) eastbound, all journeys; (b) westbound, all journeys.

- The average travel rate (**Figure 5(c)** and **Figure 5(d)**) shows rather similar results as delay for identifying the worst time periods. The average travel time index had very similar results and is not included here. The average travel rate is slightly lower for end-to-end journeys, as would be expected since they are prioritized by the signal timing. Interestingly, the performance of the evening peak does not look as good using this metric. Although the total delay is lower because volumes are lower, the travel rate is comparable to other times of the day.
- The average SOFT value is shown in **Figure 5(e)** and **Figure 5(f)**, respectively, for all journeys and end-to-end journeys. Lower values correspond to worse performance. These metrics show that eastbound/midday has the worst performance, followed by westbound/afternoon peak. The SOFT values experienced by end-to-end trips are higher, which again would be expected since this is the prioritized path. Drivers traversing the entire corridor also seem to fare worst on eastbound/midday, whereas in the afternoon peak, the two directions are about the same.

In summary, different performance measures offer different perspectives on corridor performance. The use of multiple performance measures helps reveal a more complete picture of what is occurring in the corridor. The total delay helps reveal the magnitude of the cost to drivers, while average travel rate and SOFT are able to show the relative performance across time periods when volumes are different.

## 5.2. Visualization of Corridor Performance

The aggregated performance measures offer a high-level view of the quality of progression. Visualizations of traffic flow can be used to understand better why a particular direction is performing well or poorly during a particular time of day. Using the processes described earlier, the vehicle trajectories were converted into cyclic TSDs and empirical PPDs. For illustrative purposes, two times of day are selected for a closer look: the morning peak, which tended to exhibit better performance in Figure 5, and the afternoon peak, which tended to exhibit worse performance.

Figure 6 shows cyclic TSDs for CVs traversing the entire corridor, which are easier to visualize at this scale than those containing all the journeys. In the latter case, the diagrams quickly become overwhelmed by trajectory lines, making patterns harder to identify. These figures are constructed to show two cycles next to each other to facilitate visualization. However, "Cycle 2" is a repeat "Cycle 1" image.

In Figure 5, for end-to-end journeys, during the afternoon peak, the westbound direction exhibited better performance than the eastbound direction, having a lower total delay (Figure 5(b)), a lower travel rate (Figure 5(d)), and a higher value of SOFT (Figure 5(f)), even though when all journeys were considered, the westbound direction ultimately had a higher delay (Figure 5(a)) and lower SOFT (Figure 5(e)). Similarly, the westbound direction tended to perform better for the morning peak than the eastbound direction.

These trends are also reflected in the cyclic TSDs (Figure 6). For the eastbound direction, in the morning peak (Figure 6(a)), the leading edge of the platoon is halted at N Cress Ridge (callout #1), Wacker Dr (#2), and experiences a long stop at Kennedy (#3), after which it is able to pass through the remaining intersections relatively easily (#4). In the afternoon peak (Figure 6(b)), there are more and longer stops and evidence of long queues. Evidence of queuing may be seen on the approach to NW Arterial (#5), Cress Ridge (#6), and the segment between Kennedy and Wacker nearly approaches spillback conditions (#7). The width of the platoon between Kennedy and University is very small (#8), illustrating that the green time in this direction is likely too short for the demand. The tail end of this platoon is cut off at University Ave (#9).

In contrast, the westbound direction performs relatively well in the morning peak (Figure 6(c)), although the end of the platoon is cut off at Kennedy (#10) and at Century (#11), and one rather long stop takes place at NW Arterial (#12). In the afternoon (Figure 6(d)), the platoon has a greater chance of being interrupted at certain points along the way, such as between University and Kennedy (#13), likely because of other traffic. Traffic is also stopped again at NW Arterial (#14). However, overall, the quality of progression is better than the other direction.

While these visualizations are useful, some of the trends are difficult to understand without seeing other traffic besides the end-to-end journeys. The addition of the other journeys to the diagram might provide further insight, but as mentioned earlier, these may inundate the diagrams with data and make it difficult to view individual trends. One option would be to select a smaller time period to visualize, while another is to employ the empirical PPD to show the same data in the form of cyclic distributions.

Figure 7 and Figure 8 show empirical PPDs for the morning peak and evening peak, respectively. For each direction, two PPDs are prepared: one for all vehicles and one for end-to-end vehicles only. Both diagrams use a color scale that shows the number of observations associated with each pixel. The redder the pixel, the greater the concentration of observations, meaning that there is higher flow over time in that area and greater density in that space during that time. Blue areas represent light traffic, and white shows where no CVs were counted. The intermediate colors, mostly yellow, tend to show platoons. These diagrams illustrate the size, number, relative strength, and dispersion of platoons as they move through the corridor. Queuing is also evident, shown by greater concentrations of vehicles in regions close to an intersection, transforming from diagonal motion to horizontal stopping. These areas tend to exhibit the heaviest concentration of observations.

For the morning peak, the end-to-end journey PPDs (Figure 7(a) and Figure 7(c)) exhibit similar trends as seen in Figure 6, which is not surprising since they use the same data. Overall, the westbound direction (Figure 7(c)) appears to perform better than the eastbound direction (Figure 7(a)). The addition of the other traffic (Figure 7(b) and Figure 7(d)) presents a slightly different picture, however. When all journeys are included in the PPD, we can see that most of the westbound intersections exhibit some queueing (Figure 7(d)), and the performance starts to look rather similar to the eastbound direction (Figure 7(b)). The two PPDs do not appear to differ substantially from each other as much as when only end-to-end journeys were included. In addition, some new features appear that were not visible before. In the eastbound direction, a secondary platoon appears between Kennedy and University (callout #15), while in the westbound direction, several secondary platoons are evident on several segments, such as between University and Kennedy (#16), and on the next link (#17), and so on.

The end-to-end journey PPDs for the afternoon peak (Figure 8(a) and Figure 8(c)) again show similar trends as the cyclic TSDs. The addition of the other journeys (Figure 8(b) and Figure 8(d)) increases the amount of data considerably for this time of day, and the two directions of travel again show less different from each other. Here, we can also see features emerge when all journeys are included in the PPD. Small secondary and tertiary platoons seem to be visible between Cress Ridge and Century (#18 and #19) in the eastbound direction. The link between Kennedy and University appears empty when looking only at end-to-end journeys (#20) but appears moderately busy when the other traffic is included (#21). In the westbound direction, a secondary platoon becomes visible between Century and Cress Ridge (#22), and the apparent empty space between NW Arterial and US 20 (#23) becomes full (#24).

### 5.3. Hourly Corridor Level Performance

The speed heat map visualizes the average speed for each subsegment along the corridor, enabling the analyst to observe locations and times of day where queuing and other sources of interference with traffic flow occur. Figure 9(a) and Figure 9(b) show the average hourly speed values for the eastbound and westbound directions, respectively, for all journeys. The color scales associated with this diagram indicate the average speed values computed for each 100-foot-long cell using Equation (10). Lower average speed is shown by a redder coloration, while faster speeds are greener. As previously identified in cyclic TSDs and PPDs (Figure 7 and Figure 8), the westbound direction (Figure 9(b)) exhibits better performance than the eastbound direction (Figure 9(a)). However, these figures reveal some additional details. Figure 9(a) shows that the corridor experiences the most queuing in the eastbound direction between 11 AM and 2 PM, as shown by the greater extent of slow speeds extending from University Avenue to Century Drive. In contrast, in the westbound direction, the greatest extent of queuing occurs between 2 PM and 5 PM between JFK and University Avenue. While the present study aggregated one month of weekday trajectory data, analysts could potentially adjust the data selection parameters to focus on a particular day of the week. In the future, as sample rates increase, it will likely become more feasible to examine a specific day or view the performance in near real-time.

## 6. Summary of Findings, Conclusion, and Future Work

This paper introduces a series of tools that utilize connected vehicle (CV) data to evaluate and visualize the performance of signalized arterial corridors, using the US 20 corridor in Dubuque, Iowa, as a case study. CV data was procured and processed to transform latitude/longitude coordinates into relative linear distances along road segments. An origin-destination (O-D) analysis was conducted to determine traffic flow patterns between entry and exit points (excluding driveways). A series of CV trajectory-based performance measures were then proposed for evaluating the quality of progression, including total delay, travel rate, travel time index, and SOFT. SOFT is the most recently proposed metric and is based on a Fourier transform of speed measurements to understand the degree to which the vehicle speeds are disturbed. Next, three visualizations were proposed to allow an analyst to see details of the performance: a cyclic time-space diagram (TSD), an empirical platoon progression diagram (PPD), and corridor speed heat maps. Results were shown for US 20 using data from weekdays in October 2021.

The key findings are that the high-level performance measures showed differing pictures depending on whether only end-to-end journeys were included or all journeys. In some cases, one direction of travel would appear better than the other depending on which O-D paths were included, for example. Visualizations using the TSD and PPD offered more detail. The TSDs revealed a great deal of information about the major street through traffic.

The PPDs were able to show the same overall trends but also supported the addition of other O-D paths, which revealed new features not visible when considering only major streets through traffic. The speed heat map provides an additional view of average speeds by hour and location, showing the extent of queues and other sources of inference with traffic flow. Altogether, the visualizations offer a way to understand why certain times of day perform better or worse than others.

This study presented performance metrics and visualization tools which consider CV trajectories traversing the corridor rather than aggregating or stacking individual signals' performance metrics. These visualizations could be used to identify locations where engineering solutions such as timing plan adjustments, geometric improvements, or access management may be desirable and to evaluate the outcomes. Whereas many of the current methods in ATSPM systems, for example, are focused on each intersection as a basic spatial unit, trajectory data can expand the spatial dimension to explore traffic behavior on segments with an unprecedented level of detail.

The rapid progress in infrastructure-free sensing technology has significantly broadened the scope of CV trajectory-based traffic management and monitoring systems, thereby enhancing the significance of this study. Future work will attempt to infer signal timing from the CV trajectory data, introduce information about the signal timing into the TSD and PPD visualizations, and improve on previous methods of offset optimization by better considering the spatial dimension of the problem.

## Acknowledgements

This work was supported in part by the Iowa Department of Transportation. CV data used in this study were obtained from Wejo, Ltd. The contents of this manuscript represent the views of the authors, who are responsible for the facts and the correctness of the data presented herein, and do not necessarily reflect the official views or policies of any organizations.

## Conflicts of Interest

The authors declare no conflicts of interest regarding the publication of this paper.